# Coupling dynamical and collisional evolution of small bodies II : Forming the Kuiper Belt, the Scattered Disk and the Oort Cloud

Sébastien Charnoz*

*AIM, Université Paris 7, Sap/CEA*

*91191 Gif Sur Yvette Cedex France*

&

Alessandro Morbidelli

*OCA, B.P. 4229, 06304 Nice Cedex 4, France*

\* To whom correspondence should be addressed: charnoz@cea.fr

Page = 47

Figures = 11

Tables=1

Proposed running head: Forming the Kuiper Belt and the Oort Cloud

Direct Correspondence to :

Sébastien CHARNOZ

Sap / CEA Saclay

L'Orme Les Merisiers, Bat. 709

91191 Gif sur Yvette Cedex

FRANCE

*charnoz@cea.fr*

**Abstract**

The Oort Cloud, the Kuiper Belt and the Scattered Disk are dynamically distinct populations of small bodies evolving in the outer regions of the Solar System. Whereas their collisional activity is now quiet, gravitational interactions with giant planets may have shaped these populations both dynamically and collisionally during their formation. Using a hybrid approach (Charnoz & Morbidelli 2003), the present paper tries to couple the primordial collisional and dynamical evolution of these three populations in a self-consistent way. A critical parameter is the primordial size-distribution. We show that the initial planetesimal size distribution that allows an effective mass depletion of the Kuiper belt by collisional grinding, would decimate also the population of comet-size bodies that end in the Oort Cloud and, in particular, in the Scattered Disk. As a consequence, the Scattered Disk and the Oort Cloud would be too anemic, by a factor 20 to 100, relative to the estimates achieved from the observation of the fluxes of long period and Jupiter family comets, respectively. For these two reservoirs to have a sufficient number of comets, the initial size distribution in the planetesimal disk had to be such that the mass depletion by collisional erosion of in the Kuiper belt was negligible. Consequently, the current mass deficit of the Kuiper belt, needs to be explained by dynamical mechanisms.

## 1. Introduction

It is generally believed that the main comet reservoirs, the Kuiper Belt, the Oort Cloud and the Scattered Disk (denoted as KB, OC and SD hereafter), have been formed and shaped after the formation of the giant planets. Their orbital structure as well as their total mass provide important clues to unveil the properties of the primordial planetesimal disk. Different scenarios have been proposed (see below) to explain the formation of Kuiper Belt, the Scattered Disk or the Oort Cloud. However, all these populations originated from the same planetesimal disk (although possibly from different portions of it), so that every scenario of Kuiper Belt formation should also be investigated with respect to the formation of the Oort Cloud and Scattered Disk (and vice-versa), for what concerns the size distribution, the total number of bodies etc. In particular, different models of the Kuiper Belt evolution have provided different estimates of the initial size-distribution of planetesimals in the outer disk, whose consistency with the formation of the OC and the SD needs to be tested. This is precisely the goal of this paper.

Since the discovery of the first Kuiper Belt Object (Jewitt and Luu, 1993), an intensive observational and modelling effort has been made. It is now well accepted that the current Kuiper Belt presents a deficit of mass relative to its primordial content. The current mass of the Kuiper belt is estimated to be 0.01 to 0.1 $M_\oplus$ (see for example : Berstein et al. 2004, Gladman et al. 2001, Petit et al. 2005), whereas the estimated initial mass in the 40-50 AU region is about 10-30 $M_\oplus$ (Stern and Colwell, 1997; Kenyon and Bromley, 2004; see also Morbidelli & Brown 2004 for a review).

The mechanisms proposed to explain this mass deficit of the Kuiper Belt can be grouped in two broad categories, each of which implies a different initial size distribution.

(a) *Collisional griding over the age of the Solar System* (Stern & Colwell 1997, Davis & Farinella 1997, Kenyon & Broomley 2004). In this category of models, the primordial disk extended to the present location of the KB, and some dynamical excitation induced by giant planets triggered a collisional cascade that eroded the primordial population to its present state. The collisional grinding scenario requires that the population of big objects (r >100 km) was never larger than the today's population (bodies of such size cannot be destroyed by collisions; Davis and Farinella, 1997), and that the missing mass was entirely carried by small bodies, that are obviously easier to fragment. Quantitatively, the primordial size-distribution at the big size end had to be the same as the current one, precisely $dN/dr \propto r^q$ with q ~ -4.5  (Gladman et al., 2001; Berstein et al. 2004; Petit et al. 2005) culminating at 1-2 Pluto size bodies[1]. For the total mass to be of the order of 15 $M_\oplus$, this steep size distribution had to continue down to meter-size bodies below which it turned to a shallower equilibrium-like distribution with q~ -3.5 (Dohnanyi 1969).

---

[1] *It is not really clear how many Pluto-size bodies existed in the primordial Kuiper belt. Among known objects, 4 could be categorised as `Pluto-sized': Pluto, 2003 UB313, 2003 EL61 and 2005 FY9. However, they are in resonance, in the scattered disk or in the classical belt at large inclination so that, according to*

(b)*Dynamical depletion during the primordial sculpting phase* (Petit et al. 1999; Nagasawa & Ida, 2000, Petit & Mousis 2004). In these models, various dynamical effects (scattering by giant planets or by embedded planetary embryos, excitation of orbital eccentricities due to secular resonance sweeping) ejected most of the bodies from the primordial belt. Alternatively, it has been proposed (Levison and Morbidelli, 2003) that the original planetesimal disk was truncated at ~30-35 AU and that a small fraction of the objects was transported outward and implanted into the originally empty KB during the evolution of Neptune, following various dynamical paths (Gomes, 2003; Levison and Morbidelli, 2003 ; Gomes et al., 2004). In these models, the final small mass of the KB is related to the low probability that particles had to remain in the belt (in the dynamical depletion models) or to be captured in it (in the push-out models). Dynamical processes being size-independent, this family of models requires that the initial size distribution in the disk was the same as that observed today in the KB, but multiplied by a constant big factor (between 100 and 300, corresponding to the current mass deficit factor). In other words, the initial distribution had to have an exponent q~ -4.5 for bodies larger than 100 km , and q=-3.5 for smaller bodies, as suggested by observations (Bernstein et al., 2004). The initial number of Pluto-like bodies had to be about a few hundreds. In such a distribution, the bulk of the mass is contained in big bodies, that are obviously more difficult to break. So these scenarios imply that collisional erosion was ineffective (however, this was never been tested with collisional grinding calculations, but as we will see, the present paper will confirm this assumption).

---

*some dynamical models -see for instance Gomes, 2003 - they might have been placed onto their current orbits from elsewhere, rather than formed in the primordial belt.*

As we have seen, each category of models requires a specific initial size distribution in order to have the potential of explaining the mass deficit of the KB. So, how can we decide which initial size-distribution is more realistic and discriminate between these categories of primordial evolution models?

Then, it seems necessary to broaden the problem in order to bring into consideration additional constrains. A natural extension is to consider also the formation of the Oort Cloud and the Scattered Disk. As we said at the beginning, the Kuiper Belt, the Oort Cloud, and the Scattered Disk are populations issued from the same planetesimal disk, although from two partially distinct locations of it. Indeed, the Oort Cloud is believed to have formed from the planetesimals initially in the 5-40 AU region (but mostly between 15-40 AU), and have been scattered by the giant planets on very distant and elliptical orbits, whose perihelion distance was then lifted by galactic perturbations (see Dones et al. 2004 for a review). The Scattered Disk is made of the objects scattered by Neptune during this process that never reached the Oort Cloud, and survived on scattered orbits up to the present time (Duncan and Levison, 1997; Morbidelli et al., 2004; Dones et al., 2004). Thus, they should have originated between ~20 and 35 AU, because that is the region most affected by Neptune. Conversely, the Kuiper Belt should have formed locally at 40-50 AU, or could have been transported outward from 30-35 AU during planet migration (Gomes 2003, Levison and Morbidelli, 2003). If the OC, the SD and the KB formed from the same disk, a similar size distribution should characterise the primordial Kuiper Belt and the progenitor of the Oort Cloud and Scattered Disk populations. Of course, some difference can be possible owing to the somewhat different initial heliocentric distances. However, these differences should be moderate, given that the originally heliocentric distances differed by a factor of 2 -3 at most. We will come back to this in the discussion of the paper.

The size distribution of the progenitor of the Oort Cloud and Scattered disk populations can be constrained by the requirement that the resulting OC and SD contain a sufficient number of comet-sized bodies. Observations of new long period comets imply the current population of OC comets with $H_{10}<11$ (radius $R>R_{10}$, with $R_{10}$ in the range 500m - Bailey and Stagg, 1988- to 1.2 Km -Weissman, 1996 ) is between $2 \times 10^{11}$ (Francis, 2005) and $10^{12}$ (Wigert and Tremaine, 1999). Francis's number may be more reliable, because it is based on the detection rate of new long period comets by the LINEAR survey. Previous estimates, conversely, were based on the flux of new long period comets modeled by Everhart (1967) from the rate of discoveries by amateurs. The estimate of the Oort Cloud population by Francis is also close to the estimate of Heisler (1990) and we assume, conservatively, that this figure concerns comets with R>500m. These observational constrains deal only with the visible part of the Oort Cloud, namely the Outer Oort Cloud, which population might be comparable to the Inner Oort Cloud (Dones et al. 2005). So the total Oort Cloud's population (inner+outer) may be roughly twice the Francis's number. In conclusion, we will use $4 \times 10^{11}$ as a standard value in this paper for the total population of the Oort Cloud.

Similarly, observations of Jupiter family comets imply that the current population of SD comets with $H_{10}<9$ is $\sim 5 \times 10^{8}$ (Duncan and Levison, 1997, Rickman 2005). Again, assessing the nuclear size of these comets is problematic. Levison et al. (2000) estimated that the ratio S between the number of comets with R>500 m and those with $H_{10}$<9 is between 1 and 9, with a mean of 5. Accordingly, the number of comets with R> 500m in the SD would be ~2.5x10$^9$. So, we could consider that in the current state of our knowledge, a reasonable estimate of R>500m bodies in the scattered disk is about 10$^9$.

So, the issue that we address in this paper is how many Oort-Cloud and Scattered Disk bodies can be expected (assuming the size distributions required in the two previously described categories of models for the origin of the mass deficit of the KB) ?. This is a non-trivial issue, because collisional erosion can affect the planetesimal population during its ejection towards the Oort Cloud (Stern & Weissman 2001) or their evolution in the Scattered Disk. The magnitude of the collisional erosion depends critically on the size distribution and on the dynamical history (Charnoz & Morbidelli 2003) and therefore it requires a careful evaluation. The residual number of comets surviving collisional erosion during the OC and SD formation process may be too small, especially in the case of a disk with a size distribution like that suggested by models of collisional grinding of the KB. Conversely, the opposite may be true for an initial size distribution like that supposed by the scenario of dynamical depletion of the KB.

In summary, in this work, we would like to discriminate between the scenarios of Kuiper Belt mass depletion, by looking at the implications that they would have on the total number of comets in the Oort Cloud and Scattered Disk. In the following section of the paper we present our approach to the problem. We explain how we set up a dynamical model for the excitation of the KB and OC/SD formation, and how we couple the collisional evolution to the dynamical evolution of each individual particle. In section 3, we present the results obtained assuming the two size distributions required by the collisional grinding scenario and the dynamical depletion scenario for the KB. The discussions and conclusions are collected in section 4.

## 2. Description of the method

### 2.1 The algorithm

In order to fulfill our goals, we need to compute the evolution of the size distribution of a disk of planetesimals undergoing a complex and rapid dynamical evolution under the gravitational influence of the giant planets. The method employed here is essentially the same as in Charnoz and Morbidelli, 2003 (CM03 in the following text). It requires two steps :

(1) First, a dynamical simulation is run. In this case, our simulation reproduces the "classical" scenario of OC/SD formation and KB sculpting. The giant planets are assumed to migrate through a disk of planetesimals, initially distributed between 5 and 50 AU, modelled with 6000 test particles. Following Malhotra (1995), the planets are forced to migrate by a quantity $\Delta a$ (equal to -0.2AU for Jupiter, 0.8AU for Saturn, 3AU for Uranus and 7 AU for Neptune) and approach their current orbits exponentially as $a(t)=a_{inf}$ $-\Delta a \times e^{-t/4My}$, where $a_{inf}$ is the current semi-major axis (Fig.1). Doing so, they eject on very elliptical orbits almost all of the planetesimals initially located up to ~40 AU- a fraction of which will be stored in the Oort Cloud due to the effect of the galactic tide and another fraction will remain in the Scattered Disk for the age of the Solar System. In our simulation we do not model the galactic tide, but we simply assume that a fraction, f, of the particles that acquire an elliptic orbit with aphelion distance larger than 1000 AU are effectively stored in the cloud. We take f=9% for the present work (see section 2.2 for details), as a representative value of previous estimates (Fernandez and Brunini, 2000, Dones et al. 2004, Brasser et al., 2006). Conversely, most of the population initially in the 40-50 AU range is in not ejected. About 70% of them remain in the Kuiper belt at the end of the 4Gy time-span of the simulation. However, their orbital distribution is strongly affected by the sweeping of the mean motion resonances with Neptune. Thus, in this scenario there is no effective dynamical depletion of the KB, and the current mass deficit needs to be explained by collisional grinding. The orbital elements of all particle are output every $10^3$ years for an accurate description of their dynamics, and stored in a file.

(2) Second, a collisional evolution simulation is run, by associating a size distribution to each of the 6000 test particles. The idea is that each test particle is a tracer of a small sub-set of the total population. From the orbital elements of the test particles, collision frequencies are computed for all pairs of test particles in the disk. Due to the very long dynamical integration, it is not possible to do direct detection of collisions between test particles, as in CM03, because this would have been too time consuming. Instead, the collision rate between all pairs of test-particles on given orbits is computed at each output timestep, using the classic Öpik approach (Opik , 1951; Wetherill, 1967), which assumes that the distributions of the angular elements of all the particles are uniform. This approach is slightly less accurate, but it is faster, so that we can compute collisional evolution over billions years, rather than over $10^5$ years only as in CM03. However, several tests have shown that we can reproduce the results of CM03 using the Opik approach for the computation of collision probabilities. Given that the collisional evolution is computed after the dynamical simulation, it is evident that in our model collisions cannot affect the dynamics (see CM03 for a discussion of this limitation).

The size-distributions transported by each test particle are represented by arrays containing the total number of objects in each mass bin, as explained in CM03. So a total of 6000 size distributions (one per test particle) are evolved conjointly to track with accuracy the collisional evolution of the disk, along with the pre-computed dynamical evolution. A logarithmic grid with 85 mass bins is used, with a factor of two in mass between adjacent bins. The corresponding sizes range from $7\times10^{-3}$ m to 1900 km in radius. The fragmentation model is the same as in CM03, and includes both catastrophic break-up and cratering. It requires a prescription for the material strength (denoted as $Q^*$), defined as the minimum amount of energy per unit volume to break a body such that the biggest fragment contains half of the parent-body's mass. Different models of $Q^*$ have been proposed in the past, using different assumptions. We choose the $Q^*$ function published in Benz & Asphaug (1999), which is a standard for icy bodies in the modern literature (see discussion in section 4). In this model, the weakest bodies are around 100m in size with $Q^*=6\ 10^5$ erg/cm$^3$. Some very low values of $Q^*$, smaller than $10^4$ erg/cm$^3$, has been postulated (Kenyon & Bromley 2004), in order to reproduce the low mass of the present Kuiper Belt after 4.5 billions years of evolution. However, we prefer not to use such values as they are not related with known properties of any icy material.

**2.2 Dynamical classes**

The strength of our approach is that it is possible to study simultaneously the collisional evolution of populations of bodies belonging to different dynamical classes. In order to do this, it is enough to accumulate the information carried by the size distributions associated with the test particles as those that belong to the considered class (in our case, the Kuiper Belt, the Scattered Disk and the Oort Cloud).

We define "Kuiper Belt" particles with semi-major axes between 40 and 50 AU and perihelion distance larger than 35 AU. As a consequence of the migration of Neptune that we have imposed, the number of KB particles evolved with time, from around 1000 particles to about 700. This large number of test particles ensures a very good statistical representation of the different sub-populations of the Kuiper Belt (resonant, classical, hot objects etc, see Morbidelli and Brown, 2004, for a review).

We define "Scattered Disk" particles which survive at the end of the simulation on elliptic orbits with a> 50 AU. We choose this threshold because the initial particle disk was truncated at 50 AU, so that particles with final semi major axes larger than this limit have been effectively transported outwards by the scattering action of Neptune. We find 7 particles in the Scattered Disk out of a total of 6000 integrated particles, ~5,000 of which were on unstable orbits.

We define "Oort Cloud" particles those which are eventually scattered beyond a distance of 1000 AU on elliptic orbit (~2500 particles in our simulation, about half of the total number of unstable particles, the remaining ones being ejected on hyperbolic orbits or colliding with the Sun). In reality only a fraction of the particles on these elliptic orbits are stored in the Oort Cloud region by several large scale perturbing effects (galactic tides, passing stars etc., see Dones et al. 2004 for a review). As we do not include these effects, in the present simulation we simply assume that only a fraction of 9% of the particles on elliptic orbits with aphelion distance larger than 1,000 AU end up in the Oort Cloud. Taking into account that about half of our particles reach this dynamical state, this fraction is in agreement with modern simulations of Oort Cloud formation (Dones et al., 2004), which found that 5% of the active particles in the disk are stored in the Oort Cloud up to the present time. Thus, to compute the size distribution in the Oort Cloud, we cumulate the size distributions transported by all particles that reach such elliptic orbits and multiply the result by 0.09, to account for the low efficiency of implantation in the OC. Models assuming that the solar system was embedded in a stellar cluster (Fernandez and Brunini, 2000; Brasser et al., 2006) find a larger efficiency of OC trapping, of 9 - 15% of the initial planetesimal disk, namely 2 - 3 times higher than in Dones et al. (2004). However, most of the trapped bodies end up in a very tight inner Oort Cloud, which is not a direct source of comets. It is unclear which fraction of these can be transferred into the outer Oort Cloud (the source of long period comets for which the estimate of $4 \times 10^{11}$ comets applies) by passing stars and giant molecular clouds over the age of the solar system. Thus, we think that 15% is a generous upper bound of the real fraction of disk planetesimals that end up in the active portion of the Oort Cloud.

**2.3 Initial size distributions**

The starting size distribution is the critical parameter of the problem, and very different collisional history may happen depending on this choice. The starting size distributions that we consider have to satisfy initially the following properties:

(a) The total mass is about 15 $M_\oplus$ in each 10 AU-wide heliocentric annulus of the planetesimal disk. This value comes from the estimated primordial mass of the KB (which is about 10 AU wide) and assuming an initial surface density decaying as the inverse of the heliocentric distance, consistent with the most recent disk models (see for instance Hueso and Guillot, 2005).

(b) The initial distribution is a broken power-law. Bodies smaller than an initial break-radius ($R_{to}$) have a shallow size distribution with differential size exponent -3.5, and big bodies with radii larger than $R_{to}$ have a steeper size distribution, with q< -3.5. A broken power-law size distribution is predicted by formation models, the two slopes representing regimes dominated by fragmentation or accretion. The observations of the current KB population suggest that at the big-size range q ~ -4.5 (Gladman et al., 2001; Petit et al., 2006). Therefore, in the present work we use the following prescription :

$dN/dr \propto r^{-3.5}$ for r< $R_{to}$

$dN/dr \propto r^{-4.5}$ for r> $R_{to}$ (Eq. 1)

(c) The biggest bodies are of sizes comparable to those of Pluto, Triton, or 2003 $UB_{313}$ namely about 1000 km in radius. Again, this is inspired by observations of the current KB population, and assumes implicitly that there has been no dynamical depletion, so that the largest bodies that we see today are also the largest bodies that existed initially. If dynamical depletion has been an effective mechanism, we cannot rule-out that bodies larger than Pluto existed in the past in the KB region. However such bodies have little influence on the initial number of comets or Pluto-like bodies that we assume, because they lay in the steep portion of the distribution at the big-size end, whereas the total mass is mainly contained in small bodies with size of order of the turn-over radius. Thus, their putative presence would not have a big effect on our results. Consequently, we preferred not to include such hypothetical bodies. The initial number of big bodies (i.e bodies with radii ~1000 km) $N_0$ is linked to $R_{to}$ by the constraint (a) on the total mass. Thus, the only free parameter in our model size distribution is the initial break-radius $R_{to}$ and we will now always refer to it.

Numerical simulations of the accretion process suggest that initial turn-over radius $R_{to}$ should be not smaller than 1 m, and maybe around 100m (Kenyon & Luu 1999). Observations (Bernstein et al., 2004) suggest that today's turn-over radius $R_{to}$ is somewhere in the range 45-100 km. So, without additional hypotheses, the initial turn-over radius could be within broad ranges, from 1m to 100 km. Consequently, the number of Pluto-size bodies in each 10 AU-wide annulus of the disk varies from a few (corresponding to $R_{to}$ ~1m) to a few hundred (for $R_{to}$ ~100 km) (see Fig. 2). Given that we have characterised the current OC and SD populations in terms of the number of objects larger than 500m in radius, it is important to know how many of such bodies are initially present in the disk, as a function of $R_{to}$. In each 10 AU-wide annulus, this number ranges between $10^{12}$ and $10^{13}$ (Fig. 3). It reaches a maximum of ~$10^{13}$ for $R_{to}$

=500m, and it is about $10^{12}$ for $R_{to}$ ~1m or $R_{to}$ ~100 km. Bearing in mind that only a small fraction of these bodies will be implanted in the Oort Cloud due to the inefficiency of the dynamical process (Dones et al., 2004; Brasser et al., 2006) and that the disk participating in the formation of the Oort cloud is ~35 AU wide (from ~5 to 40 AU), it is evident from the beginning that collisional grinding must have been quite ineffective in order to build an Oort Cloud containing ~ $4 \times 10^{11}$ of these bodies (Stern & Weissman 2001, CM03).

Below, we consider two nominal size distributions as two representative cases. The first one has $R_{to}$ =1m, and consequently ~1 Pluto-size body in each 10 AU-wide annulus. This size distribution is that required by models that assume no dynamical depletion in the KB and that argue that the entire mass depletion of the belt was due to collisional grinding. The second size distribution has $R_{to}$ =100 km, and ~300 Pluto-size bodies. This size distribution is that required by models invoking a dynamical depletion of the Kuiper belt of about a factor of ~100-300. We choose this specific one because the total number of bodies larger than 500m in radius is the same as for the first distribution. Therefore, the difference in the final number of bodies larger than 500m that we will obtain in the OC and in the SD, assuming the first or the second distribution *will depend exclusively* on the effectiveness of the collisional grinding.

Below, we first focus on the size distribution defined above as example cases. Then, we will also study the dependence of the results on the initial break-up radius $R_{to}$. In general, we will denote as `erosional distributions' those with $R_{to}$ <1km, as most of the mass is in small, easy-to-break bodies. Conversely, we will call `non-erosional distributions' those with $R_{to}$ >10km, for which the mass is mostly carried by bodies that are difficult to disrupt.

## 3. Collisional evolution

In this section we present the collisional evolution of the Oort Cloud, Scattered Disk and Kuiper Belt populations over the age of the Solar System, for the initial size distributions defined above.

### 3.1 Initial distribution with $R_{to}$ ~ 1m

We start with our nominal distribution with a primordial turn-over radius $R_{to}$ ~1m. Figure 4 shows the evolution of the mass of the Kuiper belt in this case, over the age of the solar system. The solid curve shows the evolution of the mass, if no collisional evolution is assumed. As we have said above, about 70% of the population remains in the Kuiper belt, so that the mass drop due to the sole dynamical evolution is only 30%. Notice that most of this mass is lost starting from about 10 My, when the Kuiper belt is excited by the sweeping of the outer mean motion resonances with Neptune. However, when collisional evolution is considered, the mass drop of the Kuiper belt is much more pronounced. Dynamical steering induces a rapid collisional cascade that affects mainly small bodies, which are fragile (Fig. 5). Since they represent the majority of the mass, the net result is a rapid decrease of the total mass of the Kuiper Belt by about a factor of 50, when summing over all size bins (dashed curve in Fig. 4), thus achieving a final mass of ~0.17 $M_\oplus$. Conversely, bodies larger than 1 km suffer a total mass decrease by a factor of 2.5 only (dashed-dotted curve in Fig. 4, to be read against the scale reported on the right vertical axis). Of this mass depletion, about half is due to the dynamical depletion, discussed above.

In the resulting size distribution of the Kuiper Belt, the turn-over radius $R_{to}$ has moved from 1m to about 3 km. More specifically there is a transition region, with a somewhat shallower slope between 100m and 2km. Below 100 m the size distribution has q~-3.5 and for big bodies beyond 3 km, q~-4.5. This value of the resulting turn-over radius is in good agreement with previous numerical simulations (Kenyon & Bromley 2004) that found a break-radius in the range 1-10 km when some simple models of gravitational stirring caused by Neptune or by embedded big planetesimals is taken into account. Note however than some analytical models predict a resulting turn-over radius around 50 km (Pan and Sari, 2005; Kenyon and Broomley, 2004); however, they rely on the assumption of constant impact velocities over the age of the Solar System, which may not be very realistic. As noted by Kenyon & Bromley (2004), when a time-dependant gravitational stirring is taken into account, the resulting turn-over radius always seem smaller than derived from analytical models, that assume constant impact velocities over the age of the Solar System.

Concerning the mass depletion, our results are again in agreement with previous studies of the collisional erosion of the KB (Kenyon and Bromley, 2004; Kenyon and Luu, 1999; Pan and Sari, 2005). For example, Kenyon and Bromley (2004) found that, assuming Benz and Asphaugh (1999) Q* law, the mass of the KB can be reduced only by a factor of 10 over the age of the Solar System. For lower values of Q*, the depletion factor could be as large as 50. Here, using Benz and Asphaugh's Q* we obtain a total depletion factor of 50. However, a factor of 1/3 is given by dynamical losses (30% of the particles leave the Kuiper belt under the effect of resonance sweeping), so that the mass depletion factor that we obtain due to the sole collisional process is about 30. Thus, we get 3 times more collisional grinding than Kenyon & Bromley (2004). The difference is probably due to two reasons. First, our KB is more dynamically excited. In fact, our KB population underwent the mean motion resonance sweeping caused by Neptune's migration, whereas Kenyon and Bromley estimated the excitation of the KB from the magnitude of the pertubations raised by a fully grown, but non-migrating Neptune. Second the size distribution considered by Kenyon and Bromley had an initial turn-over radius $R_{to}$ ~50m, namely was slightly less erosive size distribution than our distribution with $R_{to}$ =1m.

Such an overall good agreement with previous works suggests that the details of the dynamical evolution are not crucial for the collisional evolution as far as the size distribution is concerned, as long as the magnitude of the orbital excitation is about the same.

In the light of these results, should we conclude that the mass deficit of the Kuiper belt can be explained by the collisional grinding process, and that the initial size distribution of planetesimals in the outer Solar System was mass-dominated by small bodies?

A first indication that this might not be correct comes from the inspection of the final KB size distribution (Fig. 5). As the figure shows, the size distribution has preserved the original steepness (q~-4.5) down to a size R~3-5 km. Observations, however, show that the current size distribution has a break-radius of 50-100 km (Bernstein et al., 2004). So, our simulation, despite giving a good result in term of total mass, provides a final size distribution that seems inconsistent with observational constraints. However, given that the Bernstein et al. detection of the turn-over radius is still challenged (see Petit et al., 2006), we cannot consider this problem as a final disproof of the collisional grinding scenario.

The collisional grinding scenario, however, runs into an even more severe problem if we turn our attention to the formation process of the Oort Cloud and of the Scattered Disk.

We first consider the Oort Cloud. The resulting OC size distribution is plotted in figure 6. In the case collisions are not taken into account, the size distribution in the Oort Cloud is the same as in the initial disk. We note that in the absence of collision, about $1.3 \times 10^{11}$ bodies larger than 500m in radius would end in the OC, which is comparable to the estimated value of $4 \times 10^{11}$ (Table 1). If collisions are taken into account, however, the final size distribution gets strongly depleted below R~ 1 km. Indeed, bodies ejected to the Oort Cloud pass through a phase during which they have orbits with large eccentricities and moderate periods, so that they can suffer frequent, high velocity impacts (Stern and Weissman, 2001; CM03). Consequently, only $1.2 \times 10^{10}$ bodies larger than 500m survive the collisional evolution (Table 1), which is about 30 times smaller than required to form the current Oort Cloud. Remember that the total population in the Oort cloud is inferred from the flux of new long period comets with $H_{10}<11$. For our results to be consistent with the estimated OC population, the comet radius corresponding to $H_{10}=11$ should have to be ~80 m in radius, which is well below the most conservative estimate (500m: Bailey and Stagg, 1988). Alternatively, one could assume that the trapping efficiency in the Oort cloud is much larger than the value that we adopted. But even considering a 3 times higher trapping efficiency in the inner Oort Cloud typical of a dense galactic environment (Brasser et al., 2006), the number of comets that we would obtain would still be a factor of 10 smaller than the $2 \times 10^{11}$ required to form only the visible part of the Oort Cloud (Francis et al., 2005). Thus, even stretching estimates as much as we can, the OC that our model produces appears too anaemic.

Another potential problem with the results of our simulation concerns the resulting size distribution of Oort cloud comets. As Fig. 6 shows, the comets stored in the OC have preserved their pristine, steep distribution for R> 2km. The information on the size distribution of new long period comets are sparse, but there is a general consensus that it is quite shallow, overall, with a cumulative size distribution index of about ~-2 (Weissman, 1996), instead of -3.5, as found here, which is directly the primordial size distribution of big bodies. In, short, despite the collisional evolution is intense, bodies larger than 1 km keep their original size distribution even in the erosive case. The very existence of gigantic long period comets as Hale-Bopp (R~25 km) suggests that the today's distribution is shallow, and that there cannot be 4 orders of magnitude of difference between the number of comets with R> 25 km and that of comets with R > 1km, unlike what is shown in Fig. 6.

We now come to the case of the Scattered Disk. The final size distribution that we obtain is illustrated in Fig. 7. It is clearly more collisionally evolved than that of the KB, with a strong turn-over radius around 10km, and a very shallow slope between 100m and 10km. This is simply due to the stronger collisional activity of the scattered disk objects, due to larger average eccentricity inducing higher impact velocities. Consequently, the number of comets with R>500m is reduced by a factor 400 relative to the non-collisional case. The total number of bodies of this size is therefore only ~ $10^7$ (see table 1), which is about 100 times smaller than the number inferred from observations of the flux of Jupiter family comets (Duncan and Levison, 1997). In addition, as a consequence of the initial size distribution that we had to adopt to have a significant erosion in the Kuiper belt, the number of bodies with R> 50 km in the scattered disk would be less than 1,000. This is in sharp contrast with observational constraints (Trujillo et al., 2001) which suggest the existence of ~ 40,000 bodies of this size.

In conclusion, the initial size distribution that allows the mass depletion of the Kuiper belt by collisional grinding implies a too efficient collisional erosion during both the Oort cloud and Scattered Disk formation process. Consequently, the resulting Oort Cloud and, particularly, the Scattered Disk would not be enough populated and the final Oort Cloud size distribution appears too steep.

3.2  Initial distribution with $R_{to}$ ~100 km

We now consider the size distribution obtained by imposing $R_{to} = 100$ km. Remember that this distribution is chosen because it has a total mass and a number of bodies with R> 500m that are the same as those of the distribution with $R_{to} = 1$m, considered before. Thus, the differences in the results with respect to the previous sub-section will highlight the role of the collisional grinding process.

In this new distribution, the mass is contained in big-bodies, which are difficult to break. So, we expect that there can be only little collisional erosion, which is favourable to create an Oort Cloud and a Scattered Disk with a large number of comets. In fact, the size distributions resulting from our simulations are much less evolved than in the previous case (compare Fig. 8 with Fig. 6 and Fig. 9 with Fig. 7). This is also visible in the resulting size distribution of the resulting Kuiper Belt (Fig. 10).

The total number of comets with R>500m stored in the comets Oort Cloud (Table 1, Fig. 8) is $7.3 \times 10^{10}$, which is only a factor of 2.6 lower than the number of comets that would be stored in absence of collisional erosion, and a factor of 6 larger than the number obtained assuming the distribution with $R_{to} = 1$m. It is still a factor of ~ 6 lower than the number of comets of comparable size estimated to be in the OC ($4 \times 10^{11}$); however, we will see in section 3.3 that other size distributions of the non-erosional class ($r_b > 10$km) will allow us to improve substantially this match. The size distribution of OC comets preserves the initial size distribution for R>2km. As the initial size distribution has a cumulative index of -2.5, the size distribution of OC comets is thus in acceptable agreement with that of the observed new comets, given the large uncertainties on the latter. So, the overall results on OC formation are quite in agreement with the image of the OC that we have from observations of long period comets.

In the Scattered Disk, the number of bodies with R>500m surviving the collisional evolution is $7.2 \times 10^8$ (see table 1 and Fig. 9). This number is in good agreement with that of the estimated population in the current Scattered Disk ($10^9$; Duncan and Levison, 1997). The number of bodies with R> 50 km is close to 100,000, again in good agreement with the estimated population (Trujillo et al., 2001). It as been pointed out (Bernstein et al. 2004) that the size distribution of big bodies (>40 km) may be a little shallower than in the rest of the disk. In our simulation we do not find any evolution of such big and resistant bodies. So this observation, if confirmed, may be a memory of initial conditions.

In the Kuiper Belt (Fig. 11) there is essentially no mass erosion, given that the bulk of the mass is in big, unbreakable, bodies. At the end of the simulation, about 11 $M_\oplus$ remain in the Kuiper Belt, about 100 times more than the current mass of the current KB.
Thus, as we said in the introduction, the assumption of a size distribution with large break-radius is viable only if one assumes that some dynamical mechanism ejected from the Kuiper Belt ~99% of the objects, in a size-independent process (or, equivalently, that implanted ~1% of the disk's planetesimals in an originally empty KB). No dynamical depletion/implantation mechanism was possible in the dynamical evolution considered in this work.

**3.3 Other size distributions**

We now consider size distributions with a turn-over radius $R_{to}$ intermediate relative to the two values considered above. As shown in Fig. 3, changing $R_{to}$ affects the number of objects with R > 500 m in the disk. The maximum number is achieved for $R_{to}$ ~ 500m to 1km.

With a disk size distribution characterised by an initial turn-over radius $R_{to}$ ~1km, the number of comets with R> 500m stored in the OC, in absence of collisional evolution, would be $2 \times 10^{12}$, a factor 15 larger than in the case of the size distribution with $R_{to}$ ~1m. However, when accounting for collisional erosion in our simulation, the number of comets is reduced to $7.4 \times 10^{10}$, only a factor 6 larger than in the case with $R_{to}$ ~1m. Thus, the number of comets stored in the OC is not linearly proportional to the number of comets that would be implanted due to the sole dynamics. This is typical of all erosive size distributions (distributions with $R_{to}$ < 1 km) due to the effectiveness of the collisional grinding process. Thus an increase in the starting number of omets in the distribution only slightly modifies the number of surviving comets due to the high efficiency of the collisional erosion.

In contrast, with a disk size distribution characterised by $R_{to}$ ~10km, the number of comets with R> 500m stored in the OC, in absence of collisional evolution, would be $7.1 \times 10^{11}$, a factor ~4 larger than in the case of the size distribution with $R_{to}$ ~100 km. When accounting for collisional erosion the number of comets is reduced to $1.2 \times 10^{11}$, which is a factor 2 larger than in the case with $R_{to}$ ~100 km. Thus, for non-erosive distributions (distributions with $R_{to} > 10$ km) the number of comets stored in the OC is roughly linearly proportional to the number of comets initially in the disk.

A number of $\sim 1.2 \times 10^{11}$ comets with $R > 500$ m stored in the OC might be considered as a reasonable reproduction of the Oort Cloud population, $4 \times 10^{11}$, given the uncertainties on the latter, on the initial mass of the disk, on the efficiency of the dynamical trapping in the OC etc. Notice that both distributions, with $R_{to} \sim 1$km and $R_{to} \sim 10$km, give this number of comets in the OC, and presumably a maximum number of OC comets about 1.5-2 times larger can be achieved with an intermediate value of $R_{to}$. So, which of these distributions should be preferred? Probably that with the largest possible value of $R_{to}$, given the size distribution of new comets seems to be shallow, and that comets with $R > 2$ km preserve their initial size distribution. Thus, distributions with an initial turn-over radius $R_{to} < 2$ km probably would give comet size distributions that are too steep with respect to what is suggested by the observations.

Notice, however, that the number of Pluto-size bodies in the disk also depends on the value of $R_{to}$ (Fig. 2). Whatever $R_{to}$ in the 1- 10 km range, the number of Plutos in each 10 AU-wide annulus of the disk (i.e. in the KB) is between 30 to 70. Consequently, given that the current KB contains only 1- 3 of these objects, all size distributions that are successful for the OC formation require a dynamical depletion scenario for the primordial KB population.

The collisional history of the Scattered Disk seems even more severe and more constraining. Indeed, owing to the large eccentricity an inclinations of this population maintained over the age of the Solar System very erosive history is found here, even in the case of $R_{to}$ ~100 km. Size distributions starting with $R_{to}$ =1m or 100 km may produce comparable number of Scattered Disk comets in the absence of any collisional evolution ($3.7 \times 10^9$ and $5.6 \times 10^9$ respectively, see table 1). Things are radically changed when collisional erosion is taken into account : only 0.2% of the first population survive , with a resulting $10^7$ bodies which is about 2 order of magnitude below the required number (about $10^9$, see introduction), whereas up to 13% survive in the case of $R_{to}$ =100 km, with a resulting population about $5 \times 10^8$ kilometer-sized bodies, which is comparable to observations. The most favourable case for erosive distribution is when $R_{to}$ ~1 km : in this case, the population of the scattered disk in the absence of collisional erosion is about $6 \times 10^{10}$ (which is the highest value we have), unfortunately, because of the efficiency of erosion, even in this very favourable case, only about $5 \times 10^7$ bodies survive, which is still about a factor 20 below requirements.

4. Discussion

In this paper we have studied the collisional evolution of a size distribution of planetesimals that are dynamically stirred or dispersed by the perturbations exerted by the giant planets. First, we have done a dynamical simulation, accounting for the 4 giant planets migrating according to the prescription in Malhotra (1995). Most planetesimals with initial semi-major axis smaller than 40 AU are unstable, and about half of them acquire orbits with large eccentricity and semi-major axis. We assume, in agreement with Dones et al. (2004), that 9% of the planetesimals on elliptic orbits with aphelion distance larger than 1,000 AU are stored in the Oort cloud by the galactic tide and the effects of stellar encounters. Conversely, 70% of the bodies with initial semi-major axis larger than 40 AU remain in the Kuiper belt for the duration of the simulation (4Gy), but acquire orbits partially excited in eccentricity and inclination. This is true, particularly, for the bodies trapped in mean motion resonances with Neptune. The final orbital distribution in the KB is qualitatively similar to the one that is observed, with the most notable exception that the hot classical belt (the sub-population of the non-resonant objects with inclinations larger than 4 degrees; Brown, 2001) is not reproduced.

Then, given the orbital histories obtained in this dynamical simulation, we have computed the evolution of the size distribution associated to each of our test particles, following the algorithm detailed in Charnoz and Morbidelli (2003). Finally, we accumulated the size distributions associated to the particles that are in the Oort Cloud, in the Scattered Disk, or in the Kuiper belt, at the end of the simulation.

We have first considered a planetesimal size distribution that allows the collisional erosion of the mass of the Kuiper belt, from 15 $M_\oplus$ down to 0.17 $M_\oplus$. This size distribution is initially steep (differential size index q=-4.5) for objects of size ranging from 1 m in radius up to Pluto-size (1 object of this size). Assuming that the same size distribution holds everywhere in the planetesimal disk, we find that only $1.2 \times 10^{10}$ comets with R> 500 m are stored in the Oort cloud, about 40 times less than the estimated current OC population. This considered size distribution is too collisionally erosive, so that only ~10% of the comets larger than 1km in diameter this can survive collisional comminution during their dynamical dispersion caused by the giant planets, in agreement with what first pointed out by Stern and Weissman (2001) studied in CM03. Similarly, in the Scattered disk, the population of objects with R>500m is reduced by collision by a factor of ~400 and the final total number of objects is about two orders of magnitude smaller than that inferred from the flux of Jupiter family comets (Duncan & Levison 1997).

Then, we have considered a size distribution carrying the same total mass and having the same initial number of planetesimals with R> 500m, but characterized by a break-radius $R_{to}$ = 100 km. In this case the number of R> 500 m comets stored in the Oort Cloud is about 7.3 $\times 10^{10}$, which is a factor of 4 better than in the previous case, but still somewhat too low (by a factor of 3 to 7). However, The Scattered disk contains about 7.2$\times 10^8$ objects with R> 500m, in fair agreement with the estimated population. The Kuiper belt, remains too massive, and the number of large KB bodies is too large. To obtain the Kuiper belt that we see, it seems necessary to invoke a dynamical mechanism capable of reducing the KB population by a factor ~100, independently of size.

The most populous Oort Clouds are obtained with size distributions with turn-over radii between 1 and 10 km. In these cases, about ~1.2$\times 10^{11}$ comets larger than 500 m in radius are stored in the Cloud, which is within a factor of 2 to 5 from the estimated population. Among these size distributions, those with the largest turn-over radius seem to reproduce better the shallow size distribution that is usually attributed to new long period comets. Also in these cases, however, the KB remains too massive and contains too many large bodies. Dynamical depletion factors of about 15-80 are required to explain the current Kuiper Belt population.

So whatever the choice of the initial size distribution (erosive with $R_{to}$ <1 km or not-erosive with $R_{to}$ >10 km) we seem to face a dilemma: when the Kuiper Belt is collisionally depleted then the OC and the SD are too anaemic, whereas when the Kuiper Belt is not eroded, we end up with reasonable values for the OC and the SD. How to get out of this dilemma ?

We revisit now the assumptions that we have made, to check if the modification of one of them could allow us to obtain results that fit simultaneously both the KB, the OC and the SD populations, without having to invoke a dynamical depletion of the Kuiper belt. Our strongest assumption was that the size distribution is the same everywhere in the planetesimal disk, and that the same total mass is present in each equal-width heliocentric annulus of the disk. This is certainly an approximation.

There are two aspects of the size distribution that we could change in our models:

i) *Heliocentric distance dependence of the turn-over radius* $R_{to}$ We have seen in section 3 that to have an effective collisional grinding of the KB the turn-over radius needs to be about 1m, whereas to obtain both the Oort Cloud and the Scattered Disk containing enough 500 m bodies requires a turn-over radius of at least 10 km.

ii) *Radial profile of the surface density of the planetesimal disk.* We have seen that, assuming a size distribution such that the collisional depletion of the mass of the KB is effective, the OC contains at the end about 40 times fewer comets with R> 500 m than the nominal Oort Cloud. We have also seen, moreover, that for an erosive size distribution, the number of comets stored in the OC does not scale linearly with the number of planetesimals initially in the disk. Thus, to enhance by a factor ~ 40 the number of comets in the OC, we should increase by a factor larger than 80 the number of planetesimals in the region of the disk swept by the giant planets. The numbers are even more extreme for the Scattered Disk population.

Is it reasonable to assume such differences in initial turn-over radius $R_{to}$ or in the total mass between the precursors of the OC and KB populations? Whereas no definitive proof may be provided, it seems quite improbable in our current understanding of Solar System formation . The planetesimals ending in the OC, the SD and in the KB come from heliocentric distances which are not very different. On the one hand, the KB bodies are supposed to be formed in the 40-50 AU region, while the OC bodies formed in the 5-40 AU region. Note that the region that is dynamically the most efficient for the implantation of planetesimals in the OC is the 20-40 AU region (Dones et al. 2004), neighbouring closely the KB. On the other hand the Scattered Disk is essentially made of objects initially in the 25-35 AU region. Thus, the formation distances of OC, SD and KB bodies differ at most by a factor ~3. It seems quite improbable that the turn-over radius $R_{to}$ could change by 3 orders of magnitude within only a factor of 3 in heliocentric distance. Similarly, it is difficult to imagine that the surface number density of planetesimals drops by a factor of 80 within 20 au only, as this would imply that the surface density declined steeper as $1/r^{.5}$, where r is the distance from the Sun. Surface densities as steep as this one have never been considered in the literature, and are probably inconsistent with the mass distribution of the giant planets (Hayashi 1981; Weidenshiling 1977).

Another assumption in our model concerns the material strength law that we adopt, which is that of Benz & Asphaug (1999, BA99 below). Indeed the material strength of KB bodies is unknown. Assuming they are made of ice, numerous expressions of the material-strength exist, coming either from analytic theories (e.g Housen & Holsapple 1999), numerical simulations (e.g BA99), or even empirical expressions (e.g : Durda et al. 1998, Colwell et al. 2005, Bottke et al. 2005) designed to explain the current distribution of the Asteroid Belt or Giant-Planet's satellites. Because of the high-computational cost of the simulations, a unique law for ice strength has been used here, namely the law for water-ice derived in Benz & Asphaug (1999). This choice was motivated because previous works seem to confirm BA99's results, at least for the case of basaltic bodies (Bottke et al. 2005). However, BA99's law is is quite `resistant' when compared to other available fragmentation laws. In particular, some studies of the early collisional grinding of the KB (e.g. : Stern & Collwell 1997, Kenyon & Bromley, 2004; Pan & Sari, 2005) require in general much weaker fragmentation laws (and even strength-less bodies in the case of Pan & Sari, 2005) to grind down the mass of the KB with collisions. We have shown here that, when using the BA99's law, the Oort Cloud and the Scattered Disk are too eroded in the collisional grinding scenario case. So the result would be even worse if a weaker fragmentation law was adopted. In this respect our results are conservative. One could still object that the material strength could be a function of the heliocentric distance at which the bodies formed, so that a resistant law as BA99 could hold for the planetesimals in the giant planets region, and a much weaker law could hold in the Kuiper belt. We think that this is quite unlikely, for the same reasons exposed above concerning the size distributions.

Another possibility to reconcile the collisional erosion of the KB with the formation of the OC and of the SD is that the OC and the SD are less populated than we assumed. The flux of new long period comets, coupled with models on the dynamical injection of new comets from the Oort Cloud, constrains the number of OC comets with total magnitude $H_{10}<11$ (Heisler, 1990; Weissman 1996; Wigert and Tremaine 1999; Francis, 2005). The nuclear radius of a comet with $H_{10}=11$ is a subject of debate. Estimates in the literature range from R~500 m (Bailey and Stagg, 1988) to 1.2 km (Weissman, 1996). In order to build an Oort Cloud consistent with observations with an initial planetesimal size distribution that allows the collisional grinding of the Kuiper belt, we would need that the typical radius of a comet with $H_{10}=11$ is 80 m. Despite the magnitude to radius conversion is badly constrained, 80m is very low compared to current estimate (around 1km, Bailey and Stagg, 1988; Weissman, 1996). We note in passing that if the nuclear radius of a $H_{10}=11$ comet were really ~1.2 km, even our best, non-erosional size distributions would give an Oort Cloud which is too object-deficient. Thus our results seem coherent with the conversion from total magnitude to nuclear-size proposed by Bailey and Stagg (1988). Similarly, to obtain a scattered disk consistent with that inferred from the flux of Jupiter family comets ($6 \times 10^8$ comets with $H_{10}<9$; Duncan and Levison, 1997) the nuclear size of a comet with $H_{10}=9$ should be ~ 100m in radius, again much smaller than ever estimated by any author.

A final approximation that also needs a comment, concerns the very nature of our approach, precisely the fact of neglecting the effects of collisions on the dynamical evolution. May this approximation has consequences on final conclusions ? Physical collisions among bodies have two effects. On the one hand, they can randomize orbit's orientations a time-scales comparable to collision time. This may increase random velocities in the system, especially near resonances, so neglecting it may lead to an underestimate of impact velocities, and in turn, of the collisional activity. However, if this is true, the Oort Cloud and the Scattered Disk would be even more devoid of comets in the erosive case, strengthening our conclusions. On the other hand, on longer time-scales, after many collisions loss of energy may lead to a circularisation of orbits (Stern & Weissman, 2001). However, this effect should not be relevant, because the time-scale of eccentricity/inclination excitation by giant planets (about a few orbital periods) is much shorter than the collisional time-scale as shown in the last section of the CM03 (see in particular fig. 12 in that paper), unless one adopts a pathological size distribution, in which all bodies are cm-size (Goldreich et al., 2004). In this case, however, one would not have comet-size bodies in the disk, to build the OC and the SD from.

## 5. Conclusion

All these results seem to show that it is difficult to reconcile the collisional grinding of the Kuiper Belt with the formation of the Oort Cloud and of the Scattered Disk, because the too efficient collisional activity kills also both the Scattered Disk and Oort Cloud populations. We note that the present study shows that only models with very little collisional activity seem able to create simultaneously a substantial Scattered Disk and Oort Cloud. Therefore, we think that the today's low mass needs to be explained in a scenario of dynamical depletion or low efficiency implantation.

**Acknowledgements**

We are in debt with H.F. Levison for his advice and inputs. We thank D. O'Brien and P. Thebault for valuable discussions and crucial suggestions and the French National Program of Planetology (PNP) for financial support. We wish also to thank our referees, Joshua Colwell and Luke Dones for their comments that improved the quality of the paper

## References

Bailey, M.E., Stagg, C.R. 1988. Cratering constraints on the inner Oort cloud - Steady-state models. *MNRAS* 235, 1-32.

Benz,W., Asphaug, E., 1999. Catastrophic disruptions revisited. *Icarus* **142**,5–20.

Bernstein, G.M., Trilling, D.E., Allen, R.L., Brown, M.E., Holman, M., and Malhotra, R., 2004, "The Size Distribution of Trans-Neptunian Bodies," *Astron. J.*, **128**, 1364-1390.

Bottke W. F., Durda D.D., Nesvorný D., Jedicke R., Morbidelli A., Vokrouhlický, D., Levison H., 2005. The fossilized size distribution of the main asteroid belt. *Icarus* **175,** 111-140

Brasser R., Duncan M. and Levison H.F. 2006. Embedded star clusters and the formation of the Oort cloud. *Icarus*, in press.

Brown M. 2001. The Inclination Distribution of the Kuiper Belt. *Astron. J.*, 121, 2804-2814.

Charnoz S., Morbidelli A., 2003. Coupling dynamical and collisional evolution of small bodies : An application of the early ejection of planetesimals from the Jupiter-Saturn region. *Icarus* **166**, 141-166

Colwell, J., Esposito, L.W., Bundy, D., 2000. Fragmentation rates of small satelites in the outer Solar System. J. Geophys. Res. **105** (E7), 17589– 17600.

Davis, D.R., Farinella, P., 1997. Collisional evolution of Edgeworth-Kuiper belt objects. *Icarus* **125**, 50-60.

Dohnanyi, J.W., 1969. Collisional model of asteroids and their debris. *J. Geophys. Res.* **74**, 2531-2554.

Dones, L., Weissman, P. R., Levison, H. F., Duncan, M. J., 2004. Oort cloud formation and dynamics. In *Comets II* (M. C. Festou, H. U. Keller, and H. A. Weaver, Eds.), University of Arizona Press, Tucson,153-174

Duncan, M. J., Levison, H. F. 1997, Scattered comet disk and the origin of Jupiter family comets, Science, 276, 1670-1672.

Durda D.D., Greenberg R., Jedicke R., 1998. Collisional Models and Scaling Laws: A New Interpretation of the Shape of the Main-Belt Asteroid Size Distribution. *Icarus* **135,** 431-440

Fernandez, J.A., Brunini, A. 2000. The build-up of a tightly bound comet cloud around an early Sun immersed in a dense Galactic environment: Numerical experiments. Icarus 145, 580-590.

Francis P.J., 2005. The demographics of long-period comets. *ApJ* **635,** 1348-1361

Gladman, B., Kavelaars J. J., Petit J.M., Morbidelli A., Holman M.J., Loredo T., 2001. The Structure of the Kuiper Belt: Size Distribution and Radial Extent. *Astron. J.* **122**, 1051-1066

Goldreich, P., Lithwick, Y., Sari, R. 2004. Final Stages of Planet Formation. Astrophysical Journal 614, 497-507.

Gomes R., 2003. Planetary science: Conveyed to the Kuiper belt. *Nature* **426,** 393-395

Gomes R.S., Morbidelli A., Levison H. F.,2004.Planetary migration in a planetesimal disk: why did Neptune stop at 30 AU? *Icarus* **170**,492-507

Hayashi, C. 1981. Structure of the solar nebula, growth and decay of magnetic fields and effects of magnetic and turbulent viscosities on the nebula. *Prog. Theor. Phys. Suppl.* **70**, 35-53

Heisler, J. 1990. Monte Carlo simulations of the Oort comet cloud. *Icarus* 88, 104-121.

Housen K.R., Holsapple K.A., 1999. Scale Effects in Strength-Dominated Collisions of Rocky Asteroids. *Icarus* **142,** 21-33

Hueso, R., Guillot, T. 2005. Evolution of protoplanetary disks: constraints from DM Tauri and GM Aurigae. *A&A* 442, 703-725.

Jewitt, D. C., Luu, J. X., 2001. Discovery of the candidate Kuiper belt object 1992 QB1. *Nature* **362**, 730-732

Kenyon S.J.,Luu J.X., 1999. Accretion in the early outer solar system. *ApJ* **526,** 465-470

Kenyon S.J., Bromley B.C., 2004. The size distribution of Kuiper Belt objects. *Astron. J.*, 128, 1916-1926

Levison, H.F., Duncan, M.J., Zahnle, K., Holman, M., Dones, L. 2000. NOTE: Planetary Impact Rates from Ecliptic Comets. *Icarus* 143, 415-420.

Levison H.F., Morbidelli A., 2003. The formation of the Kuiper belt by the outward transport of bodies during Neptune's migration. *Nature* **426,** 419-421

Malhotra R., 1995. Origin of Pluto's Orbit: Implications for the Solar System Beyond Neptune. *Astron. J.* **110,** 420-429

Morbidelli A., Brown M.A., 2004. The kuiper belt and the primordial evolution of the solar system. In *Comets II* (M. C. Festou, H. U. Keller, and H. A. Weaver, Eds.), University of Arizona Press, Tucson,175-191

Morbidelli, A., Emel'yanenko, V.V., Levison, H.F. 2004. Origin and orbital distribution of the trans-Neptunian scattered disc. *MNRAS* 355, 935-940.

Nagasawa, M., Ida, S. 2000. Sweeping Secular Resonances in the Kuiper Belt Caused by Depletion of the Solar Nebula. *Astron. J.* 120, 3311-3322.

Öpik, E.J., 1951. Collision probabilities with the planets and the distribution of interplanetary matter. *Proc. R. Irish Acad* **54(A)**, 165-99

Pan R., Sari R., 2005. Shaping the Kuiper belt size distribution by shattering large but strengthless bodies. *Icarus* **173**, 342-348

Petit J.M., Morbidelli A., Valsecchi G. B. 1999. Large Scattered Planetesimals and the Excitation of the Small Body Belts. *Icarus* **161**, 367-387

Petit J.M., Mousis O., 2004. KBO binaries : how numerous were they ?. *Icarus* **168, 409-419**

Petit, J.-M., Holman, M.J., Gladman, B.J., Kavelaars, J.J., Scholl, H., Loredo, T.J. 2006. The Kuiper Belt luminosity function from $m_R$= 22 to 25. *MNRAS* 365, 429-438.

Rickman H., 2005. Transport of comets to the Inner Solar System. *In Dynamics of populations of planetary systems* . Proceeding IAU Colloquium N°197, 2005. Z. Knezevic & A. Milani, eds.

Stern S.A., Colwell J.E., 1997. Collisional Erosion in the Primordial Edgeworth-Kuiper Belt and the Generation of the 30-50 AU Kuiper Gap. *ApJ* **490**, 879-882

Stern, S.A., Colwell, J.E., 1997. Collisional erosion in the primordial Edgeworth-Kuiper belt and the generation of the 30-50 Au Kuiper gap. *Astron. J.* **490,** 879-884.

Stern, S.A., Weissman, P.R., 2001. Rapid collisional evolution of comets during the formation of the Oort cloud. *Nature* **409**, 589-591

Trujillo, C.A., Jewitt, D.C., Luu, J.X. 2001. Properties of the Trans-Neptunian Belt: Statistics from the Canada-France-Hawaii Telescope Survey. *Astron. J.* **122**, 457-473.

Weidenschilling S.J., 1977. The distribution of mass in planetary systems and solar nebula. *Astrophys. Space Sci.* **180**, 57-70

Weissman, P.R. 1996. The Oort Cloud. ASP Conf. Ser. 107: Completing the Inventory of the Solar System **107**, 265-288.

Wetherill G. W. 1967. Collisions in the asteroid belt. *Geophys. Res.*, **72,** 2429-2444.

Wiegert P., Tremaine S., 1999. The Evolution of Long-Period Comets. *Icarus* **137**, 84-121

| $R_{to}$: Initial Turn-Over Radius | 1 m | 100 m | 1 km | 10 km | 100 km |
|---|---|---|---|---|---|
| Number of Oort Cloud comets without erosion | $1.3 \times 10^{11}$ | $8.6 \times 10^{11}$ | $2 \times 10^{12}$ | $7.1 \times 10^{11}$ | $1.9 \times 10^{11}$ |
| Number of Oort Cloud comets with erosion | $1.2 \times 10^{10}$ | $2.7 \times 10^{10}$ | $7.4 \times 10^{10}$ | $1.2 \times 10^{11}$ | $7.3 \times 10^{10}$ |
| Ratio Oort Cloud erosion/no erosion | 0.098 | 0.031 | 0.037 | 0.16 | 0.38 |
| Final Mass in Kuiper Belt ($M_\oplus$) | 0.17 | 0.73 | 1.7 | 4.1 | 11.1 |
| Number of R=500m bodies in the Scattered Disk (no erosion) | $3.7 \times 10^9$ | $2.4 \times 10^{10}$ | $5.8 \times 10^{10}$ | $2 \times 10^{10}$ | $5.6 \times 10^9$ |
| Number of R=500m bodies in the Scattered Disk (with erosion) | $1.0 \times 10^7$ | $2.8 \times 10^7$ | $4.9 \times 10^7$ | $1.1 \times 10^8$ | $7.2 \times 10^8$ |

Table 1 : Number of surviving comets (bodies with radii > 500 m) in the Oort Cloud and in the Scattered Disk, and resulting mass of he Kuiper Belt for different simulations with varying values of the turn-over radius ($R_{to}$) in the initial distribution.

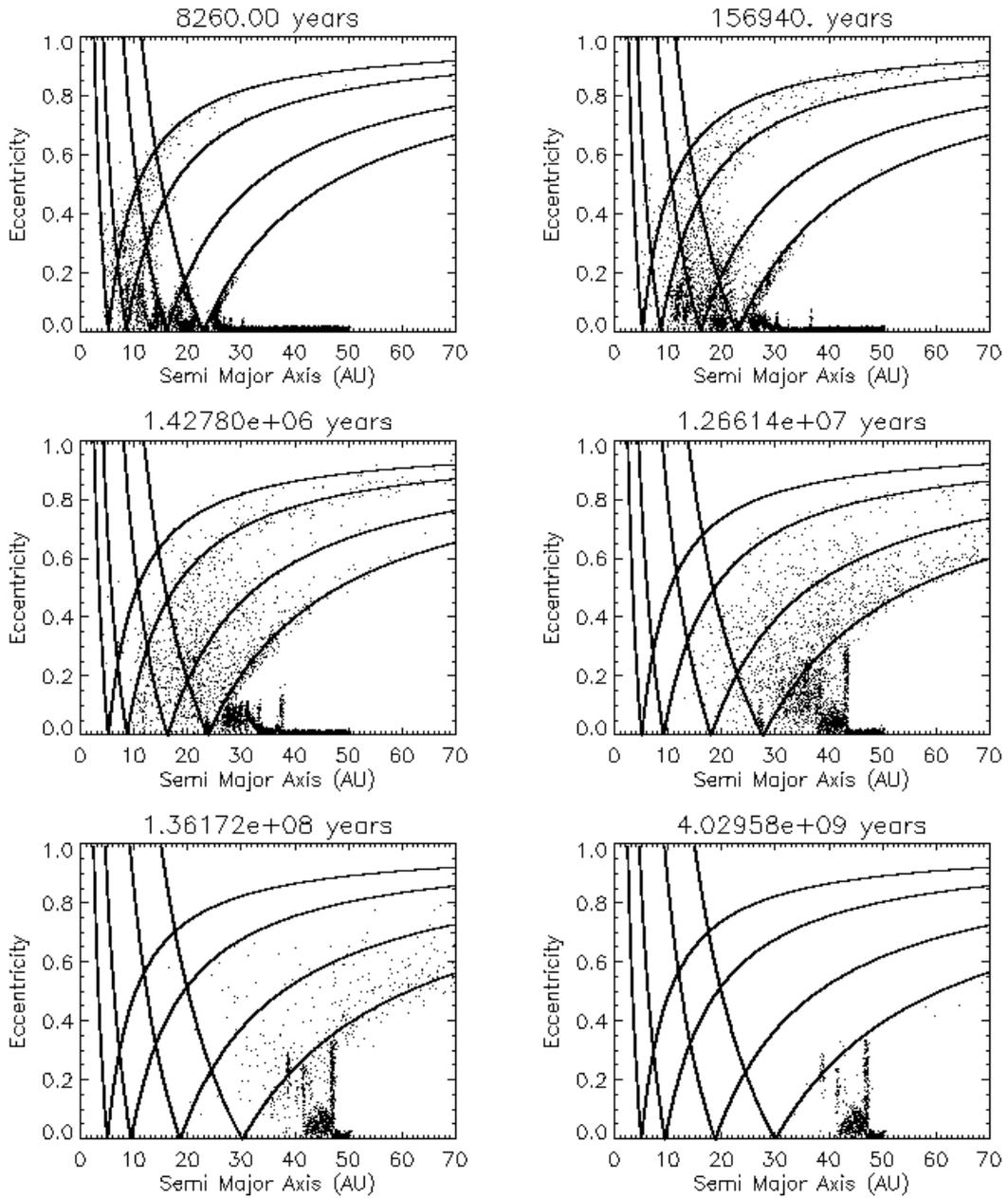

Figure 1: Dynamical evolution of the test particles under the action of the four giant planets. Giant planets migrates to their present location in $3 \: 10^7$ years. Particles above the solid lines are on planet-crossing orbits.

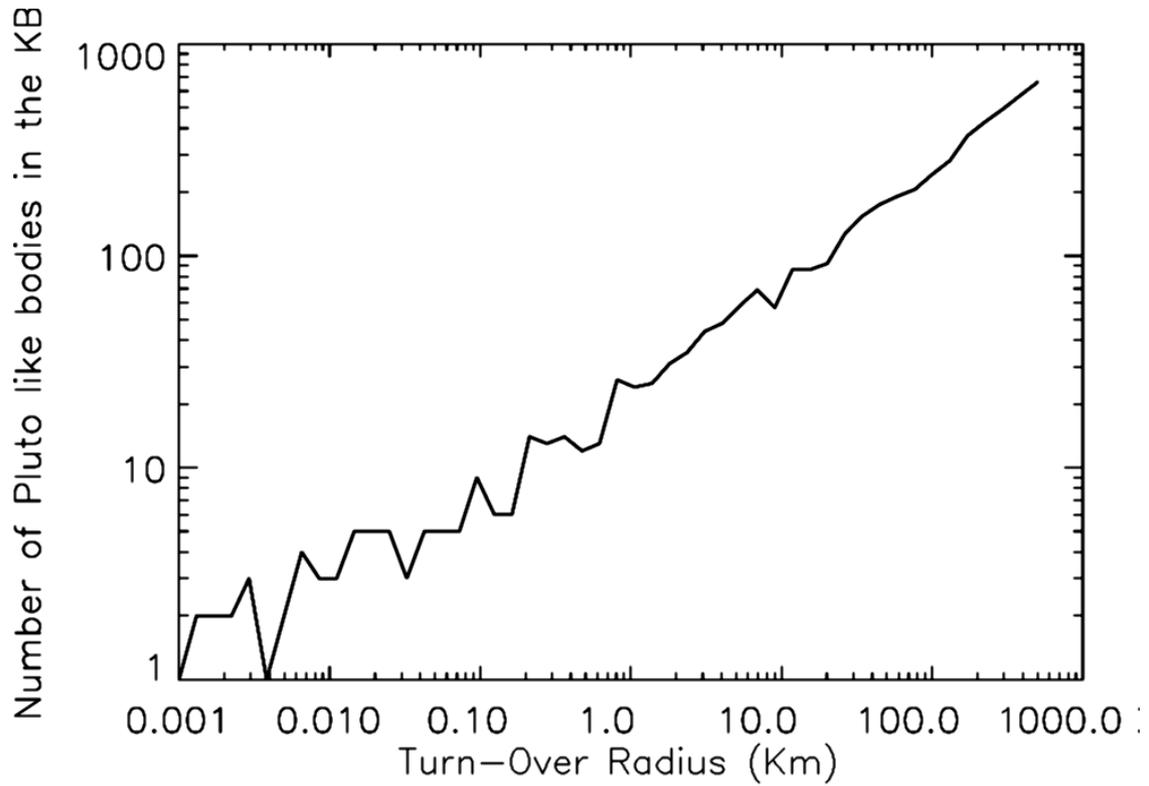

Figure 2 : Number of Pluto-sized bodies in the Kuiper Belt at the start of the simulation, as a function of the turn-over radius $R_{to}$. Noisy features comes from random generator of primordial sized distributions in the initialisation algorithm to generate integer number of bodies. We assume that the same population holds in each 10AU-wide annulus of the planetesimal disk.

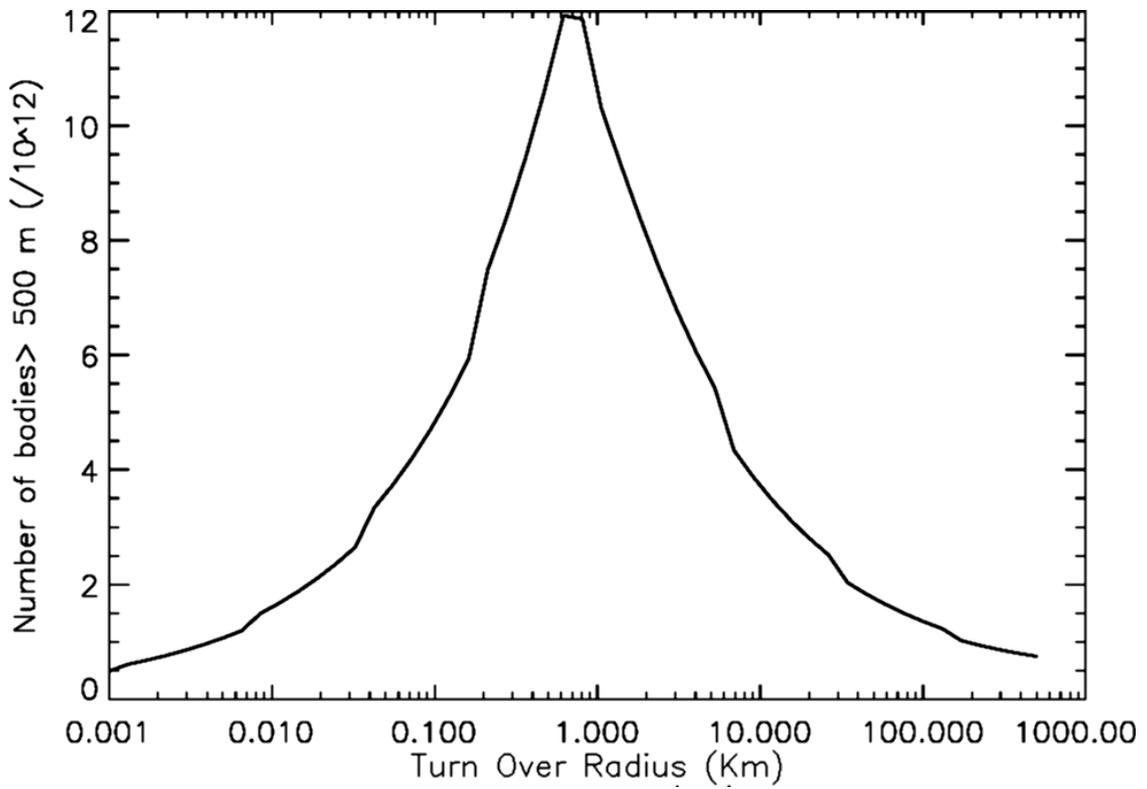

Figure 3 : Number of bodies in the Kuiper Belt with radii larger than 500 km at the start of the simulation, as a function of the inital turn-over radius $R_{to}$. We assume that the same population holds in each 10 AU-wide annulus of the planetesimal disk.

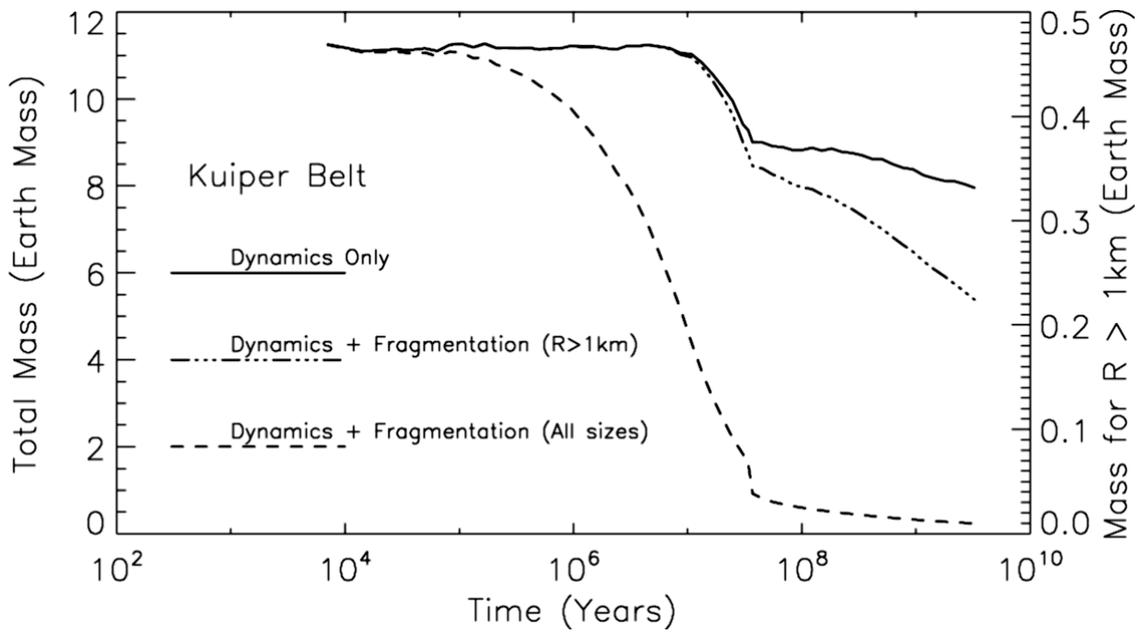

Figure 4: Mass evolution of bodies in the Kuiper Belt in the case of an initial distribution with $R_{to}$ =1m (erosional case). Solid line: mass in bodies with radius > 1km when erosion is not considered (scale on the right). Dashed-dotted line : mass in bodies > 1 km when collisional erosion is considered (scale on right). Dashed-line: mass evolution over all sizes including collisional erosion (scale on the left).

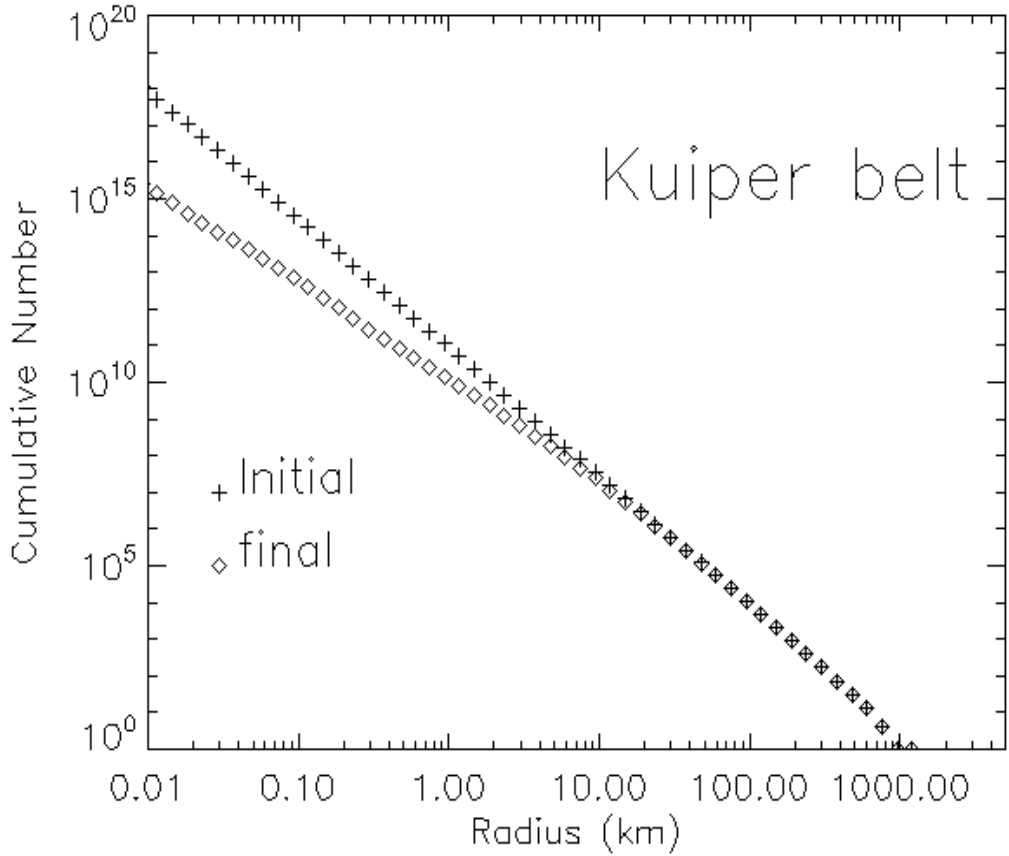

Figure 5 : Cumulative size distribution of bodies ending in the Kuiper Belt, in the case of a primordial distribution with turn-over radius $R_{to} = 1m$ (erosional case).

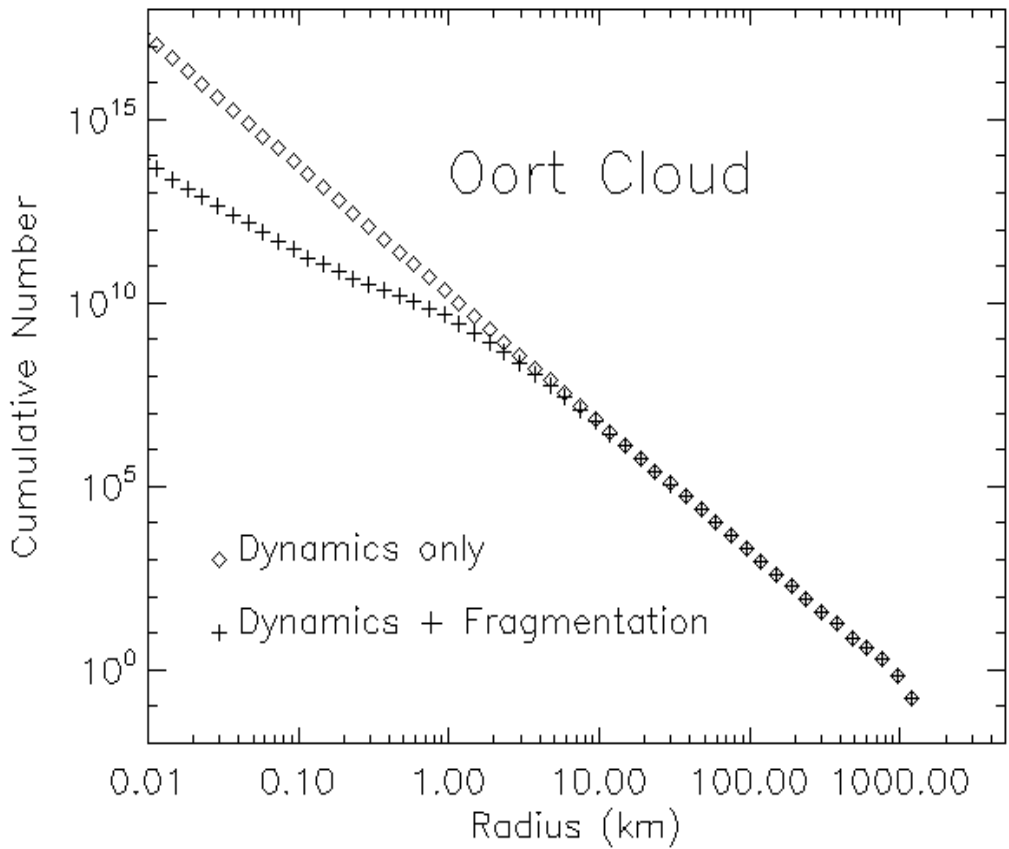

Figure 6: Cumulative size distribution for bodies ending in the Oort Cloud for an initial size distribution with turn-over radius $R_{to} = 1$ m (erosional-case). A 9% multiplicative factor is applied to take into account the efficiency of implantation in the Oort Cloud (see text). Diamonds is for the case pure dynamics i.e. without collisional erosion. Crosses stand for the case where collisional erosion is considered.

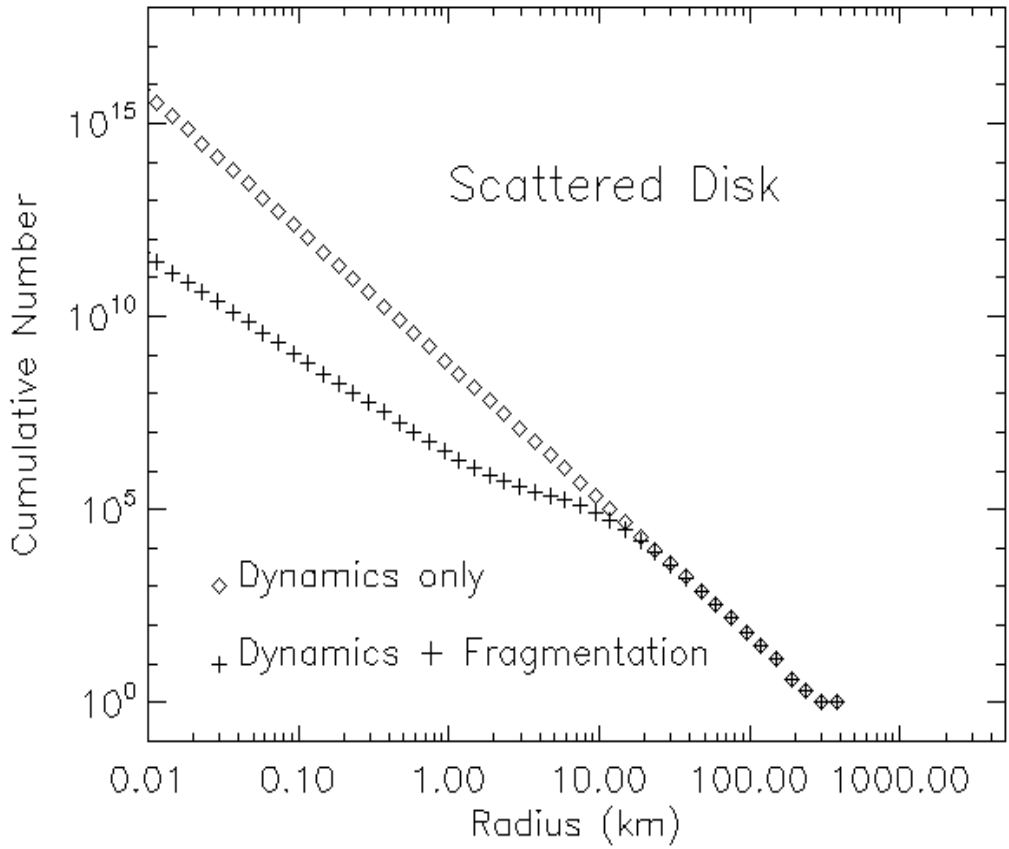

Figure 7 : Cumulative size distribution of bodies ending in the Scattered Disk, in the case of an initial distribution with turn-over radius $R_{to}$ =1m (erosional case).

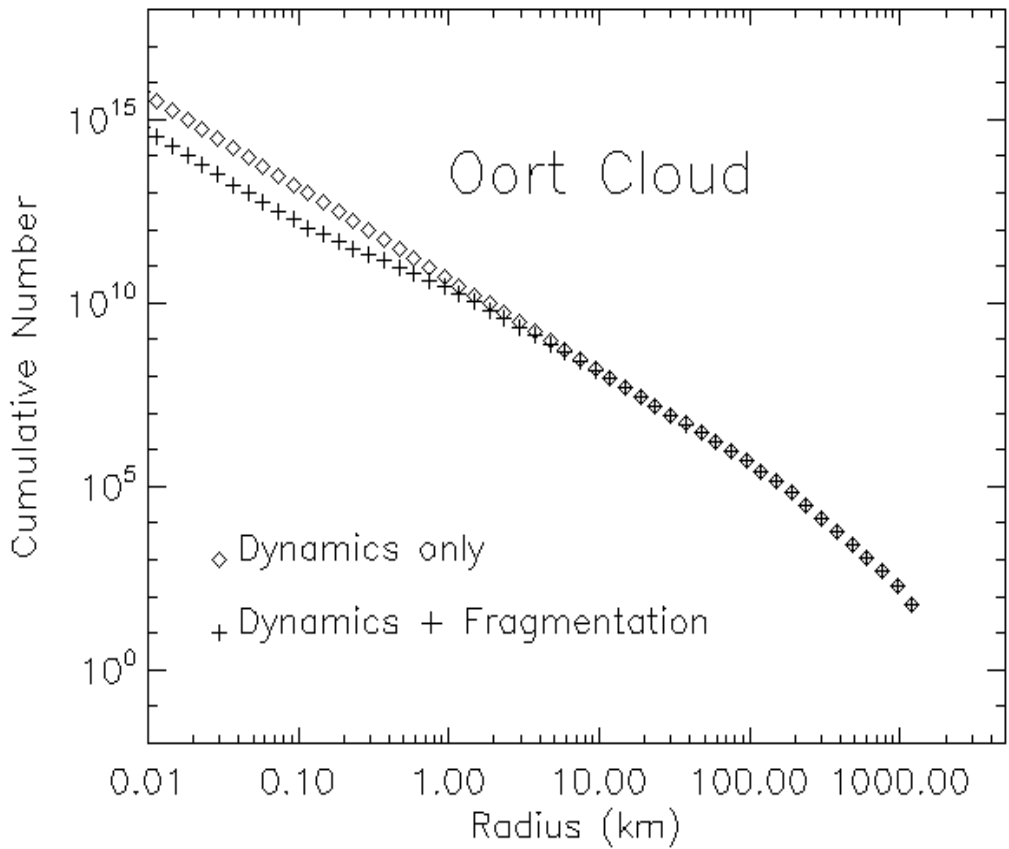

Figure 8: Cumulative size distribution for bodies ending in the Oort Cloud, for an initial distribution with turn-over radius $R_{to}$ =100 km (non erosional-case). A 9% multiplicative factor is applied to take into account the efficiency of implantation in the Oort Cloud (see text). Diamonds is for the case pure dynamics i.e. without collisional erosion. Crosses stand for the case where collisional erosion is considered.

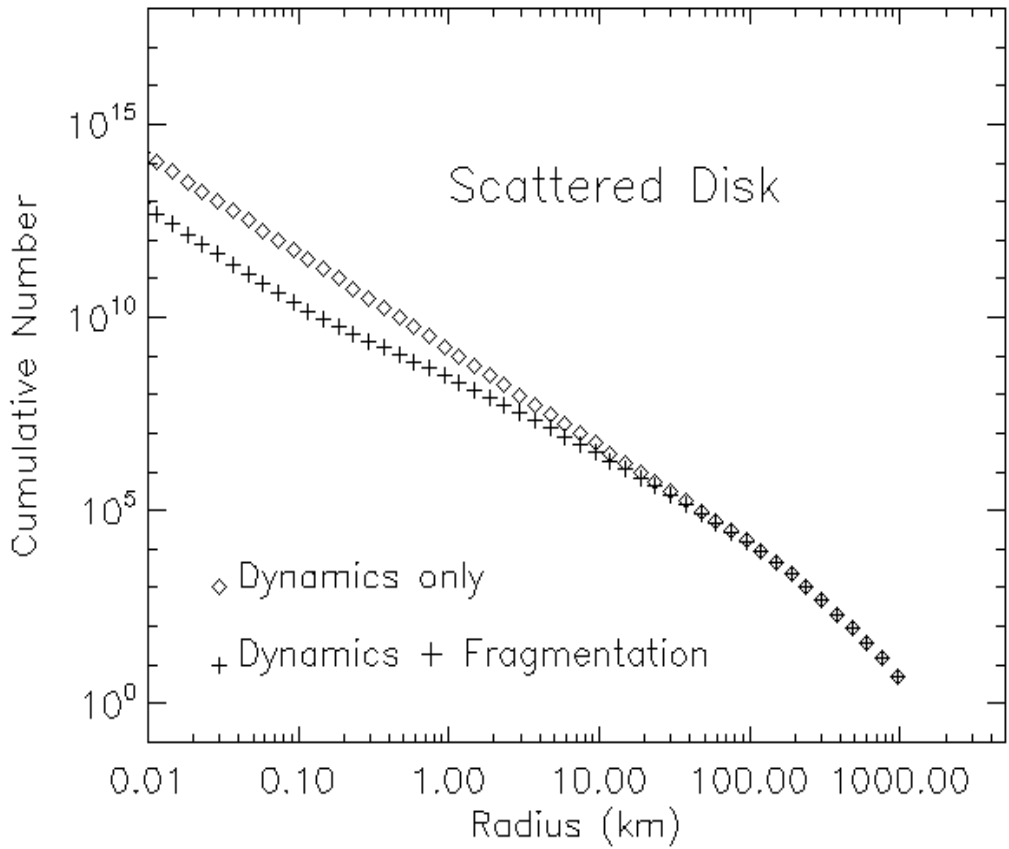

Figure 9: Cumulative size distribution of bodies ending in the scattered disk, in the case of an initial distribution with turn-over radius $R_{to}$ =100 Km (non-erosional case).

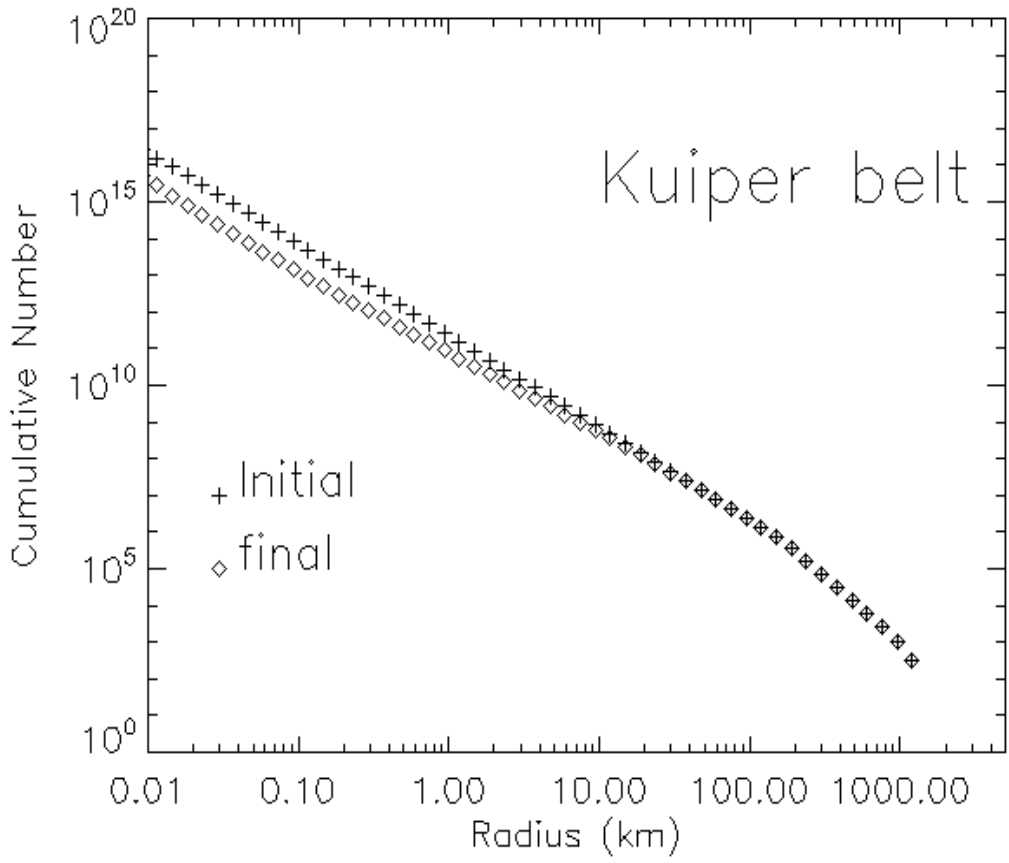

Figure 10: Cumulative size distribution of bodies ending in the Kuiper Belt, in the case of an initial distribution with turn-over radius $R_{to}$ =100 Km (non-erosional case).

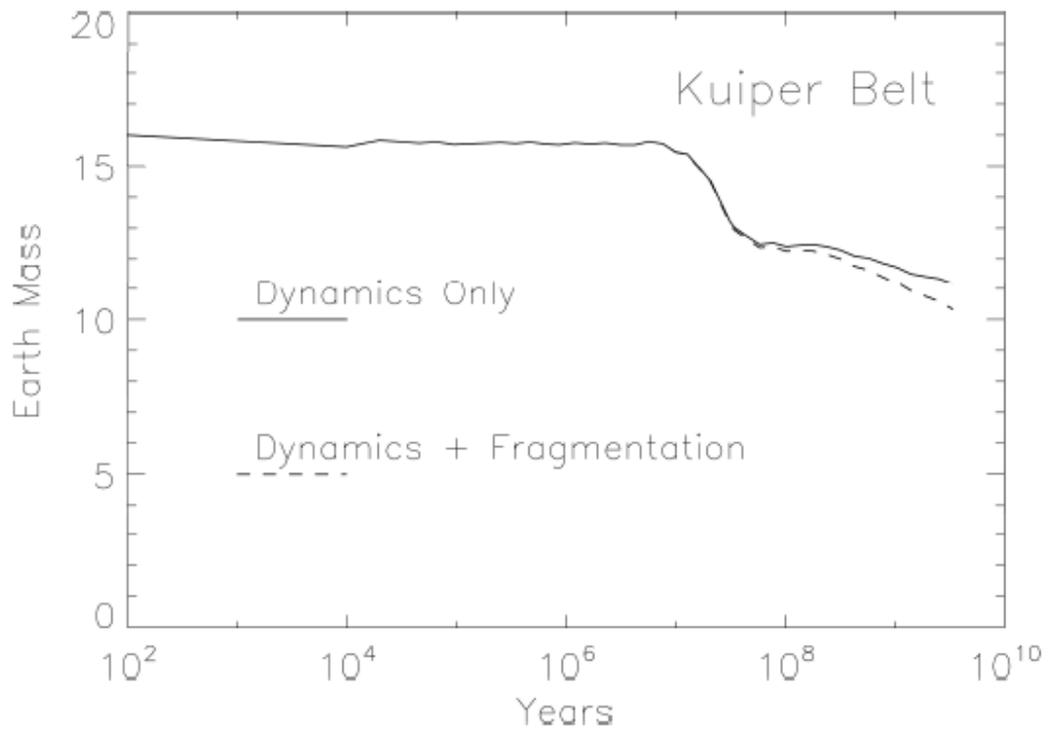

Figure11: Evolution of the total mass of the Kuiper Belt for the case of an initial distribution with turn-over radius $R_{to}$ =100 Km (non-erosional case).